# Electronic interactions of a new quatertiophene-based surfactants at liquid/gas interface.

Changwoo Bae[a], Kamatham Narayanaswamy [b], Hisham Idriss [b], Ludivine Poyac [b], Indraneel Sen [c], Sébastien Richeter[b], Sébastien Clément[b], Anne-Laure Biance [a,*], Samuel Albert[a,*], and Oriane Bonhomme [a].

We report the synthesis of a new functional molecule, a quater-tiophene based surfactant, which can both adsorb at the water / gas interface (surface active molecule) and aggregates through $\pi-\pi$ stacking interactions. We assess then the ability of this molecule to create these functionalities at interfaces. This interfacial functional aggregation, characterized here *in situ* for the first time, is probed thanks to Langmuir trough experiments and spectrometric ellipsometry. These results open some new routes for the design of new water based opto-electronic devices.

## 1 Introduction

The design of new organic materials that can assemble at the nanoscale is an essential step for the construction of the next generation of advanced functional materials for nanoscale electronics and optoelectronic devices such as OFETs, OLEDs and sensors[1,2]. As the performance of the device relies on both the optoelectronic properties and nanoscale morphology, which are inherently interconnected, precisely controlling their self-assembly at the nanoscale is of a paramount importance. To achieve this, chemists have access to a wide range of non-covalent interactions, such as hydrogen bonding, Van der Waals interactions, electrostatic forces and $\pi-\pi$ interactions that can be tuned through precise structural modifications[3,4]. Consequently, a growing focus of research on conjugated organic materials is the development of strategies to optimize their intramolecular conformation and intermolecular electronic coupling through structurally programmed non-covalent assembly[4,6].

In the self-assembly process of this specifically designed organic materials, the nature of solvent (polarity, dipole moment, protic/aprotic nature, dielectric constant) is a key parameter in view of tuning the intermolecular interactions between the organic units[7,8]. While thin films of organic materials are typically formed using organic solvent, designing molecules that can self-assemble in water while retaining their optoelectronic properties is more challenging, yet crucial[9,10].

It will indeed offer the possibility to use a sustainable and biocompatible solvent, and more interestingly be a first step for coupling ionic and more specifically protonic, and electronic transport. This coupling is generally achieved thanks to electrochemical processes at electrodes. In the specific case of interfacial processes, this coupling has also been reported in the case of electrokinetic water transport in the vicinity of Van der Waals materials such as graphene[11,12]. We explore here an alternative route, by designing new functional molecules that can at the same time assemble at interfaces and bear these electronic properties.

To this aim, we designed an amphiphilic $\pi$-conjugated molecule built upon a quaterthiophene backbone allowing the presence of $\pi$-$\pi$ stacking intermolecular interactions and three ethylene glycol chains at the periphery to ensure water solubility[13]. We measure the surface pressure of the liquid gas interface as a function of the molecular area of the molecule at interface, thanks to a Langmuir trough. We demonstrate that it tends to adsorb at interfaces and is surface active. Analysis of the curves with surface equations of state allows us to define two regimes of interaction and to evidence that aggregation occurs when the interfacial layer is dense enough. To go further, we characterize the aggregation properties of this so-called electronic surfactant directly *in situ*, when they are located at interfaces. Hence, we probe the spectroscopic properties of the molecular assembly directly at the liquid/air interface thanks to a new dedicated spectrometric ellipsometry technique. A shift of some absorption bands when the molecular area is decreased allowed us to get, for the first time, a signature of the formation of some interfacial molecular complexes compatible with electron delocalization within the molecular layer.

[a] *Universite Claude Bernard Lyon 1, CNRS, Institut Lumière Matière, UMR5306, F69100 Villeurbanne, France.*
[b] *ICGM, Univ Montpellier, CNRS, ENSCM, Montpellier, France.*
[c] *Wasabi Innovations Ltd., Sofia 1784, Bulgaria.*
\* Corresponding author: samuel.albert@univ-lyon1.fr
\* Corresponding author: anne-laure.biance@univ-lyon1.fr

## 2 Materials and Methods

### 2.1 Surfactant synthesis

#### 2.1.1 Materials

Reactions needing inert atmosphere were performed under argon using oven-dried glassware and Schlenk techniques. All anhydrous solvents and reagents were obtained from commercial suppliers and used as received without further purifications: sodium hydride (Sigma-Aldrich, > 99.5%). 1, 5'-bromo-2,2'-bithienyl-5-methanol, 5-trimethylstannyl-2,2'-bithiophene and Pd(PPh$_3$)$_4$ were prepared according to procedures described in the literature [14–17]. TLC were carried out on Merck DC Kieselgel 60 F-254 aluminium sheets and spots were visualized with UV-lamp ($\lambda$ = 254/365 nm) if necessary. Preparative purifications were performed by silica gel column chromatography (Merck 4060 M) and flash chromatography was carried out using Biotage Isolera Systems (UV-Vis 200 nm – 800 nmdetector) over silica cartridges (Sfar HC D).

#### 2.1.2 Characterization methods of the synthesised surfactant

NMR spectroscopy and MS spectrometry were performed at the Laboratoire de Mesures Physiques (LMP) of the University of Montpellier (UM).
$^1$H and $^{13}$C{$^1$H} NMR spectra were recorded on Bruker 400MHz Avance III HD and 500MHz Avance III spectrometers at 298K. Deuterated solvents CDCl$_3$ (Sigma-Aldrich, 99.8%) and DMSO-d6 (Avantor, > 99.0%) were used as received. $^1$H and $^{13}$C{$^1$H} NMR spectra were calibrated using the relative chemical shift of the residual non-deuterated solvent as an internal standard. Chemical shifts ($\delta$) are expressed in ppm. Abbreviations used for NMR spectra are as follows: s, singlet; d, doublet; t, triplet; m, multiplet.
High Resolution Mass spectra (HRMS) were recorded on a Bruker MicroTof QII instrument in positive/negative modes.

#### 2.1.3 Synthesis

**Synthesis of compound 2.** (5'-bromo-[2,2'-bithiophen]-5-yl)methanol (355mg, 1.290mmol) was placed into a two-neck 100mL round bottom flask under argon with anhydrous THF (20mL). The solution was cooled to 0 °C with an ice bath. Sodium hydride (60% dispersed in mineral oil) (33mg, 1.360mmol, 1.1eq) was added and the mixture was stirred for 1h30. Then, a solution of 1 (784mg, 1.190mmol, 0.9eq) in 15mL of anhydrous THF was added dropwise and the mixture was stirred at room temperature for 24h. The reaction was followed by TLC using CH$_2$Cl$_2$ as eluent. Once the reaction was finished, the solvent was evaporated, CH$_2$Cl$_2$ was added and the organic phase was washed with water (3 x 25mL). The organic phases were gathered, dried over MgSO$_4$ and the solvent was evaporated under reduced pressure. The resulting crude mixture was purified by flash column chromatography using gradient elution CH$_2$Cl$_2$/MeOH (98:2 to 90:10 (v:v)) leading to compound **2** as a yellow/brownish oil (0.335g, 66%).

$^1$**H NMR** (400 MHz, CDCl$_3$) $\delta$ = 6.97 (d, 1H, $^3J_{H-H}$ = 3.6Hz), 6.96 (d, 1H, $^3J_{H-H}$ = 3.8Hz), 6.89 (d, 1H, $^3J_{H-H}$ = 3.9Hz), 6.88 (dt, 1H, $^3J_{H-H}$ = 3.6Hz, $^4J_{H-H}$ = 0.6Hz), 6.58 (s, 2H), 4.63 (br. s, 2H), 4.55 (br. s, 2H), 4.14 (m, 6H), 3.85 (br. d, 2H $^3J_{H-H}$ = 4.4Hz), 3.83 (br. d, $^3J_{H-H}$ = 4.8Hz, 2H), 3.79 3.85 (br. d, 1H, $^3J_{H-H}$ = 4.4Hz), 3.78 (br. d, 1H $^3J_{H-H}$ = 5.0Hz), 3.72 (m, 6H), 3.65 (m, 12H), 3.54 (m, 6H), 3.37 (s, 9H) ppm.
$^{13}$**C NMR** (400 MHz, CDCl$_3$) $\delta$ 152.8, 140.7, 139.0, 138.0, 137.0, 133.3, 130.8, 127.4, 123.9, 123.6, 111.1, 107.4, 72.4, 72.0, 71.9, 70.9, 70.8, 70.7, 70.6, 69.8, 68.9, 66.5, 59.2.
**HRMS (ESI+):** $m/z$ calcd for C$_{37}$H$_{55}$BrO$_{13}$S$_2^+$ [M]$^+$: 851.2334Da, found: 851.2336Da.

**Synthesis of compound 3.** Compound **2** (270mg, 0.317mmol, 1eq), 2,2'-Bithiophen-5-yl(trimethyl)stannane (147mg, 0.447mmol, 1.4eq) and tetrakis(triphenylphosphine)palladium (15mg, 0.012mmol, 0.04eq) were placed into a two-neck 100mL round bottom flask and flushed with argon. Then, anhydrous toluene (25mL) was added. The solution was stirred at reflux 24 hours and the reaction was followed by TLC using CH$_2$Cl$_2$/MeOH (98:2 (v:v)) as eluent. Once the reaction was judged finished, the solvent was evaporated. CH$_2$Cl$_2$ (30mL) was added and the organic phase was washed with water (3 x 20mL). All organic phases were gathered, dried over MgSO$_4$ and the solvent was evaporated under reduced pressure. The resulting crude mixture was purified by column chromatography on silica gel using CH$_2$Cl$_2$/MeOH (98:2 to 90:10 (v:v)) as eluent leading to compound 3 as a yellow/brownish oil (260mg, 87%).

$^1$**H NMR** (400 MHz, CDCl$_3$): $\delta$ 7.23 (dd, 1H, $^3J_{H-H}$ = 5.1Hz, $^4J_{H-H}$ = 1.1Hz), 7.18 (dd, 1H, $^4J_{H-H}$ = 1.0Hz, $^3J_{H-H}$ = 3.6Hz), 7.08 (m, 2H),7.08 (br. s, 2H), 7.03 (m, 2H), 6.90 (d, 1H, $^3J_{H-H}$ = 3.6Hz), 6.59 (s, 1H), 4.65 (s, 1H), 4.47 (s, 1H), 4.15 (m, 7H), 3.85 (t, 5H, $^3J_{H-H}$ = 5.0Hz), 3.79 (t, 2H, $^3J_{H-H}$ = 5.2Hz), 3.73 (m, 7H), 3.65 (m, 14H), 3.54 (m, 7H), 3.37 (2 br s, 9H), 1.64 (s, 9H) ppm.
$^{13}$**C NMR** (400 MHz, CDCl$_3$) $\delta$ 152.6, 140.3, 139.7, 137.9, 137.5, 137.3, 136.9, 136.3, 136.2, 135.9, 135.7, 133.3, 132.1, 132.0, 131.9, 131.9, 129.0, 128.5, 128.4, 128.2, 127.9, 127.8, 127.3, 127.2, 125.3, 124.6, 124.3, 124.2, 123.7, 123.2, 107.3, 77.5, 77.4, 76.8, 72.3, 71.9, 71.7, 70.8, 70.6, 70.5, 69.7, 68.8, 66.5, 59.0, 21.4 ppm.
**HRMS (ESI+):** $m/z$ calcd for C$_{45}$H$_{60}$O$_{13}$S$_4^+$ [M$^+$]: 937.2989Da, found: 937.2994Da.
**UV-vis** (H2O): $\lambda_{max}$ (log($\varepsilon$)), 357nm (4.3), 411nm (3.8), 440nm (3.56).

### 2.2 Bulk characterization methods

#### 2.2.1 UV-Visible absorption spectroscopy

UV-Visible absorption spectra were recorded in THF, chloroform and water with a JASCO V-750 UV-Visible-NIR spectrophotometer in 10mm quartz cells (Hellma). The molar extinction coefficients ($\varepsilon$) were determined by preparing solutions of the surfactant at different concentrations in THF and water. The concentration range was chosen to remain in the linear range of the Beer–Lambert relationship ($A$ ca. 0.2–0.8).

#### 2.2.2 Dynamic Light Scattering

Dynamic light scattering (DLS) measurements were performed using a Malvern Zetasizer Nano series Nano-ZS in water. The critical micellar concentration (CMC) values were determined by preparing solutions in water at different concentrations and measuring the intensity of the scattered light. The data were visualized by plotting scattered light intensity as a function of the concentration, revealing a sharp increase at the CMC. The underlying principle is that larger particles scatter light more efficiently than smaller molecules. Accordingly, solutions of **3** at various concentrations were prepared for the measurements and showed this increase as reported in figure 3.

#### 2.2.3 Cryo-Electron Microscopy

Three microliters of suspensions prepared in deionized water were applied to glow discharged Lacey grid (Ted Pella inc.), blotted for 3s and then flash frozen in liquid ethane using a EM-GP2 (LEICA). The concentration of the amphiphilic quaterthiophene - compound **3** - was $50\,\mu mol/LM$. Before freezing, the humidity rate was stabilized at 95% at 20 °C in the chamber. Cryo-TEM observation was carried out on a JEOL 2200FS (JEOL, Europe, SAS), operating at 200kV under low-dose conditions (total dose of 20 electrons/Å$^2$) in the zero-energy-loss mode with a slit width of 20eV. Images were taken with direct detection electron K3 camera (Ametek-Gatan inc.) at a nominal magnification of 6000x, 15 000x and 30 000x with defocus ranging from 1.8 to 2.5$\mu$m.

### 2.3 Molecules at interfaces

#### 2.3.1 Deposition of the molecules at liquid-gas interfaces.

Due to the limited solubility of the surfactant molecule in water and the necessity of placing them at the water/air interface, surfactant solutions are prepared using chloroform as the solvent. A carefully measured quantity of chloroform with a density ($\rho_{chl}$) of 1.49 g.cm$^{-3}$ at a temperature of 25 °C is added to the initially powdered compound **3** using a precision balance (Sartorius). As chloroform is highly volatile, even when stored in a sealed container, the solutions are renewed before each experimental campaigns by evaporating the chloroform, measuring the weight of the resulting dried powder, and subsequently adding fresh chloroform. All experiments are conducted within one to two weeks from the date of solution preparation to maintain the consistency and reliability of the results.

The molecular layer is deposited on ultrapure water placed in a Langmuir-Blodgett trough (KSV NIMA Medium), composed of a Teflon tank with two motorized barriers positioned at the water/gas level, allowing for adjustment of the surface area accessible to the surfactants at the liquid/gas interface $A$. First, the Teflon tank is filled with ultrapure water, and a paper Wilhelmy plate is hung on a force sensor to measure the surface pressure at the interface (figure 1). Then, a given volume of the surfactant chloroform solution is dispensed cautiously thanks to a glass syringe (Hamilton) on top of the water/air interface. Great care to minimize the dissolution into the bulk water is required. The amount of volume dispensed each time was approximately $1\,\mu L$ dropwise. A small correction factor (less than 15%) was added when computed the molecular area of **3** to correct uncertainties on the deposited volume and on the surfactant concentration in chloroform. This factor is constant for a run of experiments and allow to collapse both adsorption isotherm, ellipsometry data and UV-Vis adsorbtion spectra.

#### 2.3.2 Langmuir trough experiments

To characterize the adsorption of insoluble surfactants, we measure the evolution of the surface tension with the surface concentration using the Langmuir-Blodgett trough as mentioned before. The motion of the Teflon barriers allows a decrease of the air/water interface area, noted $A$, inducing a compression of the deposited layer. The velocity of the barriers is fixed at $40\,\text{mm}\,\text{min}^{-1}$ avoiding any potential surfactant dissolution (compound 3 is hardly soluble in water). A vertical force sensor probes the surface pressure $\Pi = \gamma - \gamma_0$ of the interface, with $\gamma_0$ the surface tension of the naked water/air interface and $\gamma$ the surface tension of the interface with the adsorbed surfactant layer. Finally, the surface pressure is reported as a function of the area per surfactant molecule, or molecular area, noted $a_s$, which is $a_s = A/(nN_A)$, with $n$ the amount of surfactants deposited at the interface and $N_A$ the Avogadro number (figure 6). As explained previously, the uncertainty on $n$ was corrected by adding a corrective multiplicative factor of the order of 10%.

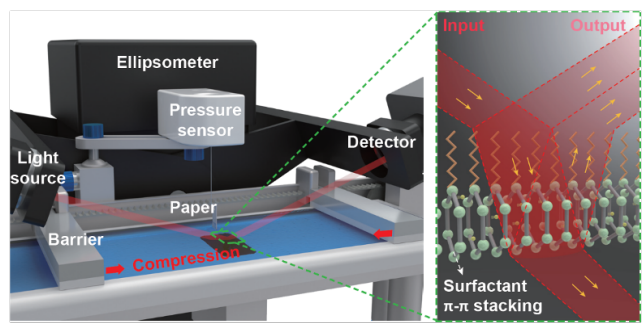

Fig. 1 Scheme of the experiments that couple spectroscopic ellipsometry and Langmuir trough characterizations.

#### 2.3.3 Spectroscopic ellipsometry on liquid interfaces

The optical properties of the surfactant layer at the air/water interface are scrutinized with spectroscopic ellipsometry. Light with different polarization is reflected on the air/water interface with a given angle of incidence (typically 65° in our case), for a wide wavelength spectrum in the range of 200 to 1000 nm. The apparatus (Woohlam M-2000 Spectroscopic Ellipsometry) measures the ratio between the reflection coefficient of the light polarized in the incident plane, noted $\rho_P$ versus the one when the light is polarized perpendicularly to the incident plane noted $\rho_S$. This ratio is a complex number that is written as:

$$\frac{\rho_P}{\rho_S} = \tan(\Psi)e^{i\Delta} = \rho + i\tilde{\rho} \qquad (1)$$

and the observables are either $\Psi$ and $\Delta$ quantifying respectively the amplitude and phase change of this ratio, or the real and imaginary components $\rho$ and $\tilde{\rho}$. The values of this complex ratio depend on the angle of incidence of the light, the thickness of the deposited layer and the refraction and absorption indexes of the layer as a function of the wavelength through a nonlinear relation. Obtaining the properties of the layer from the ellipsometry measurements requires a model and data inversion procedure. Such procedure is developed on thin solid layer deposited on a solid substrate[18,19] or in the commercial software from Woohlam. However, it is still challenging to perform it on liquid interfaces where roughness, fluctuations or heterogeneities must be considered[20–22]. Taking into account these effects is out of the scope of this study, we will only analyse here directly the evolution of the raw ellipsometry data.

The ellipsometry and Langmuir trough techniques are coupled, as depicted in 1, to discuss the evolution of the optical properties (namely $\Psi$ and $\Delta$) in relation with the adsorption isotherms (then surface pressure and surface concentration).

## 3 Surfactant synthesis and bulk characterizations

### 3.1 Surfactant Synthesis

The synthetic route to the amphiphilic quaterthiophene, noted **3**, is described in figure 2. **3** was prepared in two steps starting from precursor **1**. First, the etherification condensation reaction of precursor **1** with 5'-bromo-2,2'-bithienyl-5-methanol in THF at room temperature (RT) for 24 hours produced **2** in 66% yield. Finally, quaterthiophene (compound **3**) was synthesized by a palladium-catalyzed Stille cross-coupling reaction between 5-trimethylstannyl-2,2'-bithiophene and compound **2** and results in 87% yield. The full reaction protocol and characterizations of compound **3**, are detailed in section 2.1.3. Compounds 2 and 3 have been fully characterized by $^1$H NMR, $^{13}$C$^1$H NMR, and high-resolution ESI-MS (see Appendix 1).

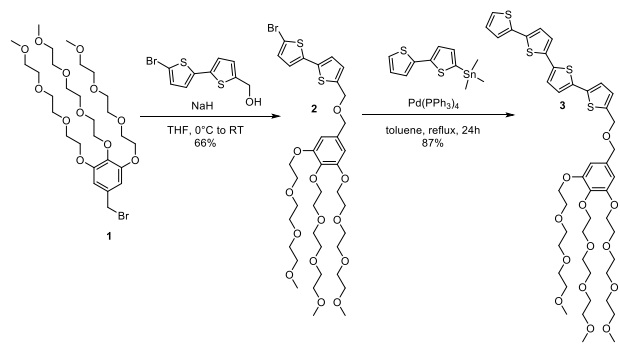

Fig. 2 Synthesis route of amphiphilic quaterthiophene 3.

### 3.2 Optical properties and aggregation behavior

To determine the critical micellar concentration (CMC) of the obtained amphiphilic quaterthiophene (compund **3**) in water, the intensity of scattered light was measured at various concentrations, as depicted in figure 3. Scattering was indeed observed, a signature of molecular aggregation of this compound in water. More quantitatively, a CMC value around 0.005mM was found for **3** revealing its strong hydrophobic character probably due to the presence of the four thiophene units. The morphology of the aggregates formed by the amphiphilic quaterthiophene **3** was then examined using cryotransmission electron microscopy (cryo-TEM). Typical morphologies obtained for samples prepared in deionized water at 0.05mM, a concentration 10 times higher than the CMC, are depicted on figure 4. Large vesicular assemblies, generally unilamellar, with diameters ranging from 60 to 600 nm, were observed, as previously reported[13] for a quaterthiophene bearing a triethylene glycol polar chain in position 2. The size distribution histogram is reported in appendix 1.

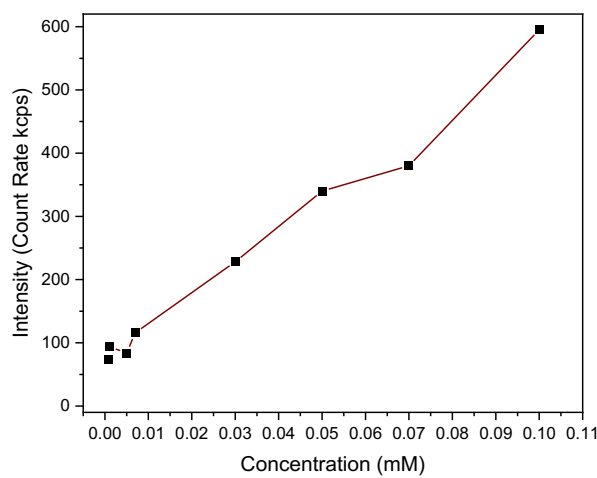

Fig. 3 Intensity of scattered light (in kilo counts per second) obtained for various concentrations of 3 prepared in deionized water.

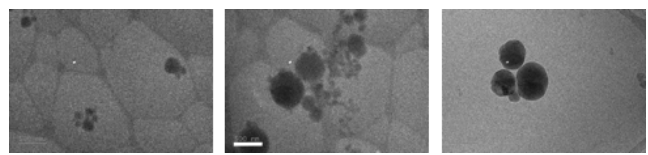

Fig. 4 Cryo-TEM of amphiphilic quaterthiophene 3 in water at 0.05 mM. Scale bar is 500 nm.

The UV-visible absorption spectra of the compound was measured in DMSO, water and chloroform as shown in figure 5. The UV-visible absorption spectrum of compound **3** in DMSO displays a broad band lacking distinct fine structures at 404 nm, indicating the presence of multiple conformers in solution. This absorption is attributed to a twisted anti conformation, deviating 30° from planarity, similar to that observed and calculated for other unsubstituted oligothiophenes[23,24]. Similar spectra were observed in chloroform. When moving to water, a blue-shifted absorption of $\lambda_{max}$ = 357 nm is noticed compared to the absorption in DMSO, which may be due to the aggregation order of the oligothiophene chromophores in water[13].

More quantitatively, the molar extinction coefficients ($\varepsilon$) are measured both in water and in chloroform. For water, the molar extinction coefficients are recorded for the several band that constitute the UV-Vis absorption spectra.

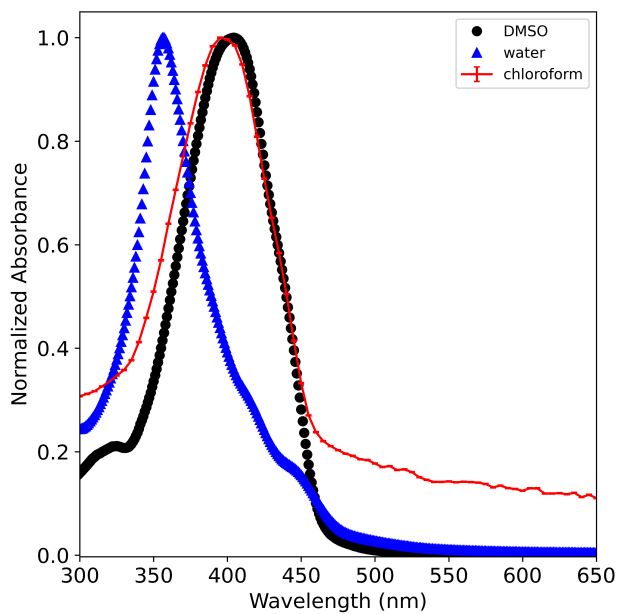

Fig. 5 UV-Visible spectra of compound 3 at 0.05mM in DMSO (black rounds), 0.05mM in water (blue triangles), and at 0.92mM in chloroform (red line).

**UV-vis** ($H_2O$): $\lambda_{max}$ ($\log(\varepsilon)$)= 357nm (4.3), 411nm (3.8), 440nm (3.56).

## 4 Aggregation properties at interfaces

### 4.1 Langmuir trough experiments

**Results** The evolution of the surface pressure as a function of molecular area shows some specific features (figure 6). First, at large molecular area, the surface pressure remains close to zero, experiencing a tiny increase during compression. Then a larger increase is observed, a signature of interactions in between the molecules at the interface, up to reaching a small plateau (or kink) for a surface pressure around 10 mN/m. Finally, a very sharp increase is observed up to 35 mN/m, followed by a rapid collapse at very large compression (small molecular area). Two main regions can then be distinguished: one at large surface area or small surface pressure (below 10 mN/m), before the kink, which will be noted as regime 1, and one at small molecular area or large surface pressure that we will note regime 2. A final regime at very large compression, where the surface pressure decreases, is a signature of the collapse of the monolayer, well-documented in the literature[25]. Compression/decompression cycles of the interfacial layer within regime 1 elicit a high degree of reproducibility, whereas some hysteresis in the isotherm appears if compression is done at surface area smaller than the kink one. A similar shape of adsorption isotherm was observed on quaterthiophene functionalized with a ethylene glycol chain[13]. A phase transition also appears but at smaller molecular area and higher surface pressure probably due to the smaller steric hindrance of the ethylene glycol groups. Moreover, the quaterthiophene surfactant synthesized in the current work appear to be more stable at interface as we can perform several compression/decompression cycles without evolution of the adsorption isotherm in the regime 1.

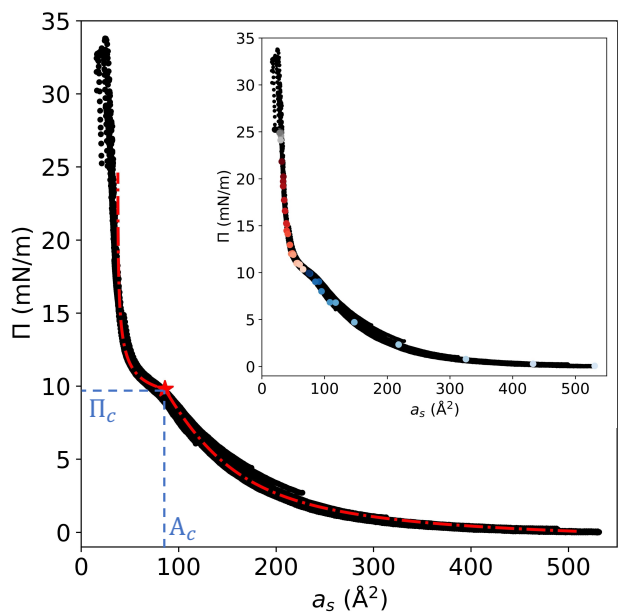

Fig. 6 Evolution of the surface pressure $\Pi$ as a function of the molecular area for KNS147 at the air/water interface at 21 °C. The evolution is obtained for compound 3 from different sets of experiments, where molecular area is corrected. Data are fitted by a combination of eqs 3 and 4 with parameters given in Table 1. Inset: Same data, with colored circles that corresponds to the point at which ellipsometry measurements were performed (figure 7).

**Discussion** To understand the surfactant behavior at interfaces in the two regimes, and to extract the relevant interaction parameters, we will use equations of state known for Langmuir monolayers[26], but including two-dimensional phase transitions[27].

For that, we first assume that regime 1 corresponds to a classical compressible monolayer, using a modified Volmer model[26,28,29]. This model is compatible with the high reproducibility of the compression/decompression cycles of the interfacial layer in this regime, consistent with the absence of strong intermolecular interactions. This model assumes that the surfactants can be considered as hard disks, with no long-range interaction. These hypotheses are softened, by introducing a 2D monolayer compressibility $\varepsilon$, which modifies the excluded area per molecule as a function of the surface pressure, as

$$\alpha = \alpha_0(1-\varepsilon\Pi), \qquad (2)$$

with $\alpha_0$ the initial excluded area per molecule, and $\Pi$ the surface pressure. An additional cohesion pressure $\Pi^*$ accounting for molecular interactions in the gaseous state is also introduced. Within these hypotheses, the equation of state then reads

$$\Pi = \frac{k_B T}{a_s - \alpha_0(1-\varepsilon\Pi)} - \Pi^*. \qquad (3)$$

When the surface area is decreased, a kink is observed, at a critical pressure noted $\Pi_c$. This kink is a signature of a phase transition, that we attribute to the formation of aggregates at interfaces[13,27].

| | | | |
|---|---|---|---|
| Regime 1 | $\alpha_0$ (Å$^2$) 99.3±0.1 | $\varepsilon$ (m/N) 53.4±3 | |
| Kink point | $A_c$ (Å$^2$) 85.9±0.5 | $\Pi_c$ (mN/m) 9.79 ±0.16 | |
| Regime 2 | $\varepsilon_a$ 0.622±0.003 | $\Pi^*$ (mN/m) -0.947±0.003 | $\Pi_a^*$ (mN/m) -40.9 ±2.9 |

Table 1 Parameters used to fit data of figure 6, using eqs. 3 and 4 and associated error estimations of these parameters

In regime 2, the Pressure - Area isotherm can be modeled by an equation of state that considers aggregation of the molecules[27], described in Appendix 2. Single surfactant molecules are then in equilibrium with aggregates of $n$ surfactants. In this case, the equation of states reads:

$$\Pi = \frac{k_B T \left(\frac{a_s}{A_c}\right)^2 \exp\left\{-\frac{2(\Pi-\Pi_c)\varepsilon_a \alpha_0}{k_B T}\right\}}{a_s - \alpha_0 \left(1 + \varepsilon_a \left(\left(\frac{a_s}{A_c}\right)^2 \exp\left\{-\frac{2(\Pi-\Pi_c)\varepsilon_a \alpha_0}{k_B T}\right\} - 1\right)\right)} - \Pi_a^*, \quad (4)$$

with $A_c$ the critical area at which the molecules tend to aggregate, $\varepsilon_a$ the normalized difference in the area occupied by $n$ single molecules and one aggregate ($\varepsilon_a > 0$ means that one aggregate occupies a smaller area than $n$ free molecules), and $\Pi_a^*$ a pressure that accounts for interactions between the aggregates at low surface area. Note that $\Pi_a^*$ and $\Pi^*$ have different values, as they account for distinct phenomena.

Such a refined model, using both eq.3 and 4 is required to fit all the data, and a satisfying agreement is achieved (see figure 6). The fitting procedure is detailed in Appendix 2. The fitting parameters are given in Table 1 and discussed below.

The molecular area occupied by one monomer $\alpha_0$ is found to be around 1 nm$^2$, showing that when they are not interacting, single molecules tend to be planar at the interface. In this dilute regime, the interactions between monomers are very small, as shown by the very small value of $\Pi^*$. The compression factor $\varepsilon$ is in the range of commonly observed factors with these types of insoluble molecules[29]. The critical area at which aggregation occurs $A_c$ is a little bit less than the area of one monomer $\alpha_0$, showing that a change of conformation of the molecule at the interface must occur before aggregation. This is confirmed by the fact that a compression factor needs to be introduced in the Volmer modified regime.

In the aggregated regime, the aggregates are more compact than the free monomers, as shown by the value of $\varepsilon_a$, which is larger than 0. This compaction is quite important ($\varepsilon_a \simeq 0.6$) demonstrating a strong reorganization of the molecules at interface, a signature of specific oriented interactions, among which electronic delocalization or so-called $\pi$-$\pi$ interactions are good candidates. The typical area per molecule in the aggregate is $\varepsilon_a/n = \alpha_0(1-\varepsilon_a) \approx 37.5$ Å$^2$ (corresponding to the area where the collapse appear on figure 6): it is slightly larger than the largest measured cross section of one thiophene group[13], probably due to the presence of the three ethylene glycol chains. It tends to confirm that the electronic surfactant does not have all its thiophene group along the interface in regime 2.

To confirm these findings at a molecular scale, we probe the surface by spectrometric ellipsometry, which gives some qualitative information of the optical response of the monolayers, and can be correlated to the absorption band of the amphiphilic quaterthiophene.

### 4.2 Spectroscopic ellipsometry
#### 4.2.1 Results

The optical properties can in turn provide indications on the aggregation and may reveal a potential electronic delocalization at the interface, which can be then discussed regarding the conductive properties of the surfactant-laden interface.

Figure 7 report the spectra of the amplitude $\Psi$ (top) and phase $\Delta$ (bottom) performed on the air/water interface covered by compound **3** at different stages of the surface compression. For a non-absorbing compounds at the interface, only variation of $\Delta$ are expected. A clear signature of the adsorption of compound **3** is observed both on $\Psi$ and $\Delta$. The amplitude of the spectra change with the surface compression but also the shape of the spectra.

#### 4.2.2 Discussion

In the presence of $\pi - \pi$ interactions and electronic delocalization at the interface, we expect a wavelength shift and/or enlargement of some of the absorption peaks of the surfactant layer. These shifts are due to the degeneracy of an energy level, where one of the degenerated energy levels is forbidden due to symmetries[30].

At Brewster angle, for surfactants that do not interact with light, the ellipticity coefficient $\tilde{\rho}$ is a signature of surfactant adsorption: it becomes more and more negative when surfactants get adsorbed at air/water interface as the hydrocarbon chain have a negative contribution in this parameter[31]. A direct link between this coefficient and the surfactant concentration is not possible without independent calibration. However, it appears that for single chain cationic surfactants, $\tilde{\rho}$ evolves linearly with the surface concentration[31]. We adopt the same approach even if we do not work at Brewster angle to discuss the ellipsometry data collected. We then first analyse the signature of surfactant adsorption far from resonance (absorption peaks) *i.e.* at 620 nm for example: the evolution of the ellipsometry ratio $\tilde{\rho}$ is plotted as a function of the surface concentration $\Gamma = 1/a_s$ on figure 8, top. At small molecular area (regime 1 on figure 6), one recovers a linear decrease of $\tilde{\rho}$ as a function of the surface concentration. This is in agreement with observations of the adsorption isotherms: the surfactants behave as independent molecules, with almost no interaction.

At the molecular area $A_c$, where there is a kink in the adsorption isotherm, the slope of the ellipticity as a function of the surface concentration changes: this is also a signature that the surfactant interaction begins. A new linear regime is observed, which is a signature of a change in the effective optical index of these interacting entities. The collapse regime identified in the adsorption isotherm is confirmed by the optical signal: there is no more evolution of the ellipticity ratio indicating that there is no more evolution of the surface concentration.

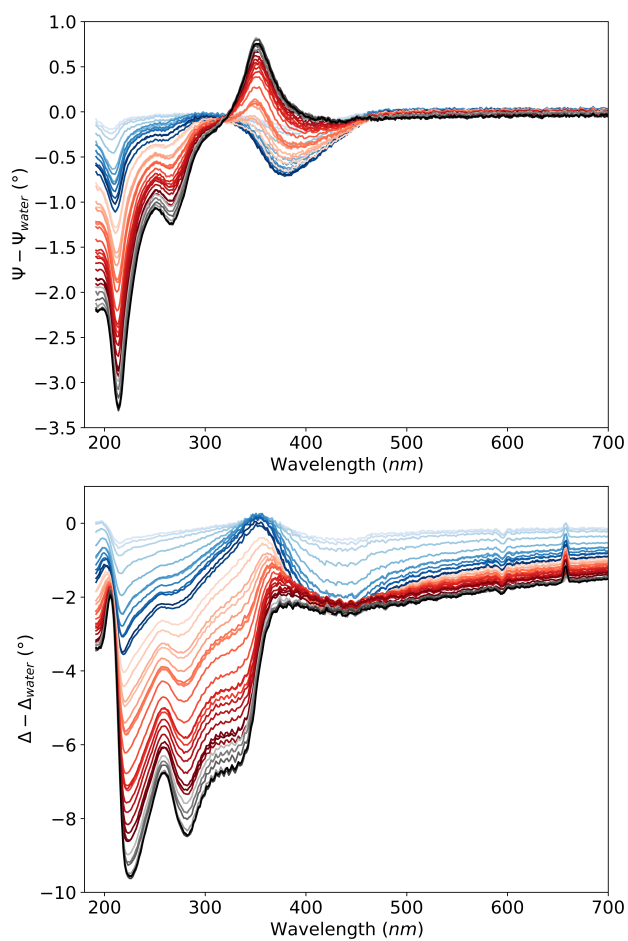

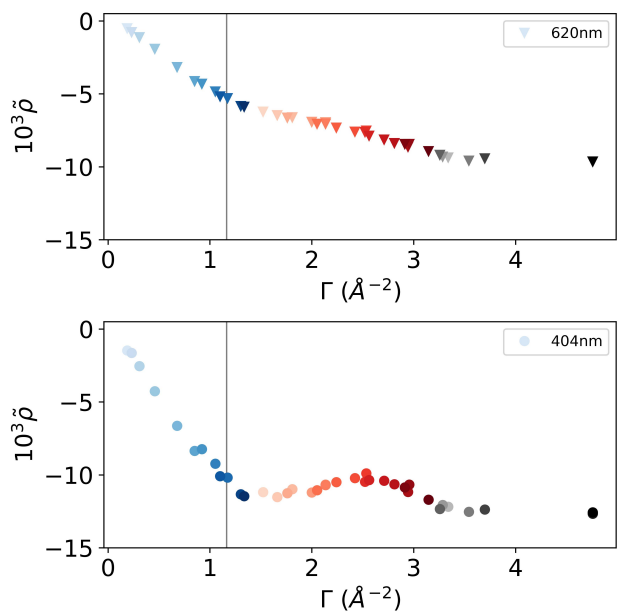

Fig. 7 Spectroscopic ellipsometry measurements performed at different stage of the surface compression. The blue curves are performed in the dilute regime (regime 1), the red ones in the aggregated regime (regime 2), and the gray ones when the monolayer collapses. The curve color corresponds to the point colors on insert of figure 6.

Fig. 8 Evolution of the ellipticity coefficient with the surface concentration respectively (top) far from adsorption band at 620 nm and (bottom) at the adsorption band in water i.e. at 404 nm. The vertical grey line corresponds is equivalent to $A_c$ (see Table 1), *i.e.* the beginning of the interacting-regime identified on the adsorption isotherm

Interestingly, the behavior close to the absorption band of the surfactant dispersed in chloroform, namely 404 nm, is very different, as illustrated in figure 8, bottom. The linear evolution of the ellipticity ratio is recovered in the dilute regime but the intermediate regime is not monotonous. Looking to the spectral signature of the ellipsometry data, one can clearly see that a hole in Ψ around 404 nm is progressively transformed in a peak at smaller wavelength, typically 350 nm very close to the peak of absorption when the surfactant is dispersed in bulk water (Figure 5). If a direct comparison between the UV-Vis absorption spectra and the ellipsometry measurements is not possible, this trend tends to highlight a shift in the absorption band under compression. Simultaneously, a new peak appears around 275 nm probably due to the appearance of a new band of adsorption, indicating once again the appearance of the aggregates.

## 5 Conclusion

In this work, we describe the synthesis of a amphiphilic conjugated molecule possessing aggregation properties at the interface. To ensure amphiphilic features, we decorated a quaterthiophene, that should allow intermolecular electronic delocalization, with three ethylene glycol chains, that should ensure its amphihilic properties. The characterization of this new compound demonstrates its bulk aggregation behavior in water. Signature of the aggregation process at the interface has also been observed thanks to adsorption isotherm, in which two distinct regimes can be identified, a signature of a 2D phase transition. A model that considers the formation of aggregates closely matches the data, and an analysis of the fitting parameters is in good agreement with the fact that specific directional interactions exist between the molecules in the aggregates, to allow compaction of the molecules. This aggregation is also confirmed by optical characterization at the interface, thanks to ellipsometric spectroscopy. Even if this technique does not allow us to get the direct spectroscopic properties of the surfactant monolayer, analysis of the signal intensity at various wavelengths shows specific transition when aggregation occurs (peak shift), which are compatible with electronic delocalization due to $\pi - \pi$ interactions.

This work offers many prospects in the design and characterization of functional surfactants at interfaces. Various similar molecules, varying thiophene backbone or the length of ethylene glycol chains can be optimized to get the best molecular assembly in the aggregates. Spectroscopic ellipsometry is a powerful tool to characterize spectroscopic properties of the interface, but also requires fundamental theoretical work to analyze the signal to get optical index of the monolayer, taking into account the presence of interface fluctuations. Finally, interfacial conductivity measurements are required to probe the real efficiency of these functional molecules.

## Acknowledgements

This work was financially supported by the European Union's Horizon 2020 Research And Innovation Program (PROGENY No. 899205) and French National Research Agency (ANR-20-CE09-0025 Soft Nanoflu). We thank the platform nanoptec at iLM and Christophe Moulin for technical support.

## Data availability

Data will be available on request.

## Notes and references

# Appendix 1: Surfactant synthesis: coumponds characterizations

$^1$H NMR spectrum, $^{13}$C$^1$H NMR spectrum and High resolution ESI-TOF (positive mode) mass spectrum of compounds 2 are reported in figures 9, 10, and 11 respectively. Similarly, $^1$H NMR spectrum, $^{13}$C$^1$H NMR spectrum and High resolution ESI-TOF (positive mode) mass spectrum of compounds 3 are reported in figures 12, 13, and 14.

# Appendix 2: Equation of state for the aggregated phase

We consider that two phases coexist at the interface in equilibrium: one phase, with single surfactant molecules, in coexistence with some 2D monodisperse aggregates of $n$ surfactant molecules. The chemical potential of each component at interface reads[32]

$$\mu_i = \mu_i^0 + k_B T \ln f_i x_i - \gamma \alpha_i, \tag{5}$$

with $f_i$ the activity coefficient, $x_i$ the molecular fraction, $\alpha_i$ the partial molecular area and $\gamma$ the surface tension. The adsorption Gibbs equation for our system then reads

$$d\Pi = \Gamma_1 d\mu_1 + \Gamma_a d\mu_a, \tag{6}$$

with $\Gamma_1$ and $\Gamma_a$ the adsorption amount of single surfactants and aggregates respectively. These equations result in the generalized Volmer's equation that reads

$$\Pi = k_B T \frac{\Gamma_1 + \Gamma_a}{1 - \Gamma_1 \alpha_0 - \Gamma_a \alpha_a} - \Pi_a^*, \tag{7}$$

with $\Pi_a^*$ a cohesion pressure. Considering that the chemical potential of the aggregates is equal to the one of the monomers (because they are at equilibrium), one can link the adsorption amount of monomers and aggregates such as $\Gamma_a = \Gamma_1 (\Gamma_1/\Gamma_T)^{n-1}$, where $\Gamma_T$ is the critical adsorption value for 2D aggregation. These quantities can be used to fit the $\Pi$-$a_s$ isotherms.

In that purpose, let's first define $A_c$ the critical area where aggregation occurs ($A_c = 1/\Gamma_c$) and $A_a = 1/\Gamma_a$ the molecular area in the aggregate regime. The molecular area of the monomer is then noted $\alpha_0$ and the molecular area of the aggregates is noted $\alpha_a$. We can define a parameter $\varepsilon_a$ that accounts for the fact that monomers in aggregates can occupy a different surface compared with the one occupied by the free monomer. It results then in $\alpha_a = n\alpha_0(1 - \varepsilon_a)$. Finally, we can assume that we have equilibrium between monomers and aggregates, so the local concentration of monomers is given by its value at aggregation $\Gamma_c$. In the case of a small numbers of aggregates, this results in $\Gamma_1 \simeq \Gamma_c$. However, when the number of aggregates is larger, the surface free of aggregates is reduced by a factor which is close to $A_c/a_s$ in our range of compression. It results then that $\Gamma_1 = \Gamma_c a_s/A_c$.

Finally, the aggregation equilibrium $n\mu_1^s = \mu_n^s$ results in

$$\frac{\Gamma_a}{\Gamma_1^n} = K_n \exp\{(\Pi n \varepsilon_a \alpha_0 / k_B T)\} \tag{8}$$

with $K_n = \exp\{((n\mu_1^0 - \mu_n^0 - \gamma_0 n \varepsilon_a \alpha_0)/k_B T)\}$ the constant of aggregation equilibrium. We then find that

$$A_T^{n-1} = K_n \exp\{(\Pi n \varepsilon_a \alpha_0 / k_B T)\}. \tag{9}$$

which can be simplified in

$$A_T(\Pi) = A_c \exp\{[(\Pi - \Pi_c)\varepsilon_a \alpha_0 / k_B T]\}, \tag{10}$$

with $\Pi_c$ and $A_c$ the pressure and area at which aggregation can occur. Combining eqs. 7 and 10 results in eq 4 of the main text.

**Fitting procedure** The experimental isotherm was fitted on the forme $a_S = f(\Pi)$ to ensure a good convergence of the fitting procedure. For that, the fitting function was defined by part. For pressure below $\Pi_c$, the critical pressure indicated on figure 6, the equation $a_S = f(\Pi)$ derived from eq.3 was used. For pressure above $\Pi_c$, the equation $a_S = f(\Pi)$ derived from eq.4 was used. The two functions are computed such as they both have a pressure equal to $\Pi_c$ when the molecular is equal to $A_c$. No limitations were applied for the different fit parameters.

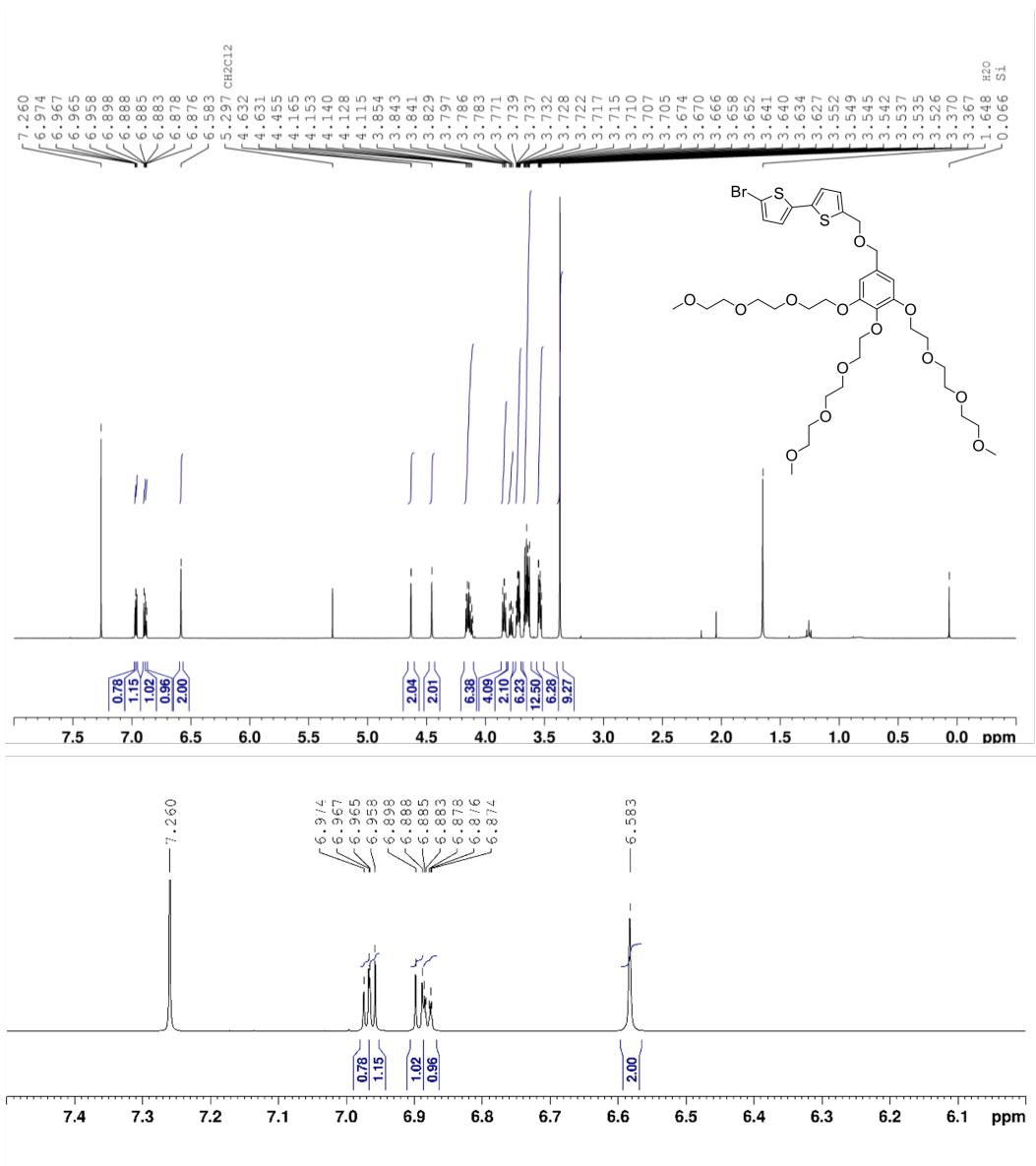

Fig. 9 Full (top) and partial (bottom) $^1$H NMR spectrum (400 MHz, CDCl$_3$) of compound 2

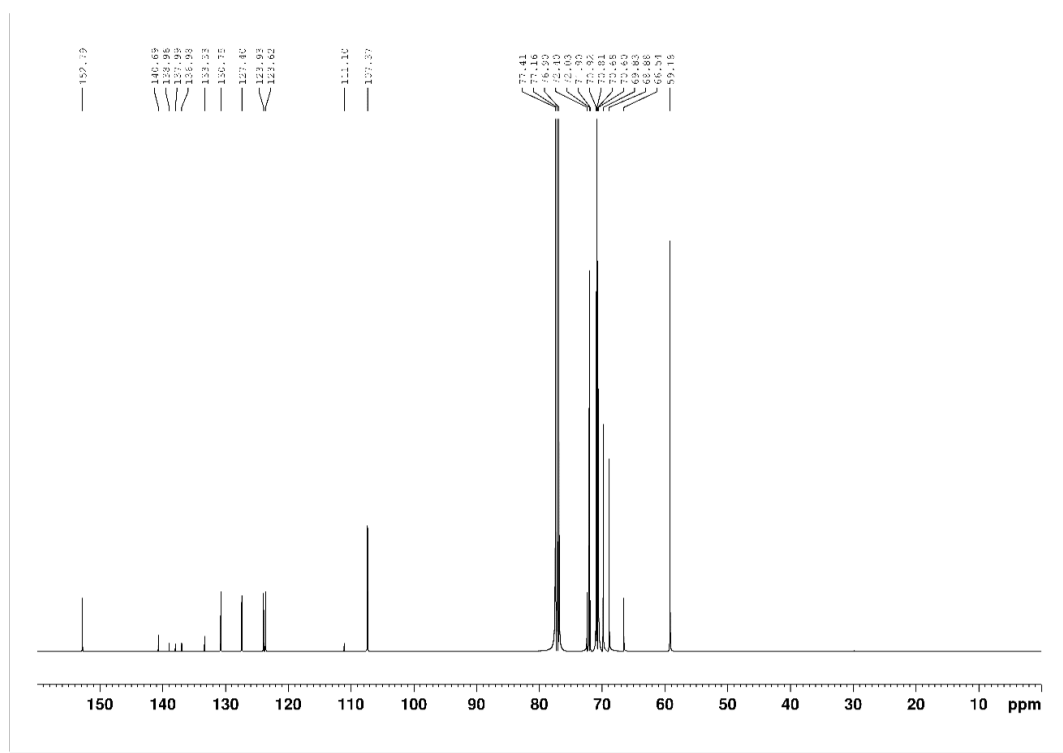

Fig. 10 $^{13}C^{1}H$ NMR spectrum (125 MHz, $CDCl_3$) of compound 2

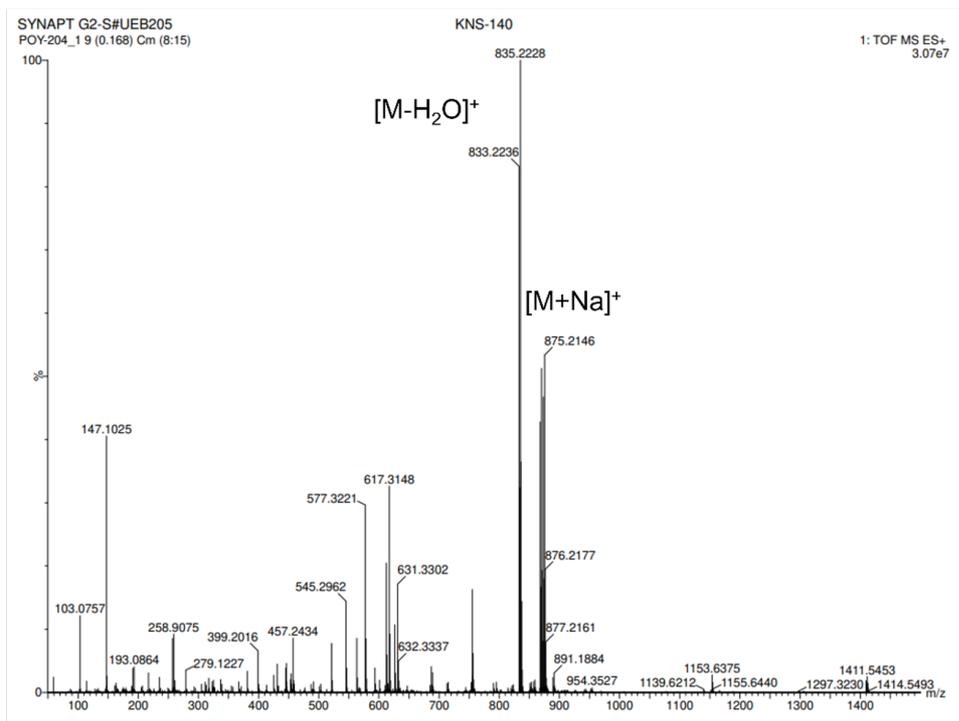

Fig. 11 High resolution ESI-TOF (positive mode) mass spectrum of compound 2

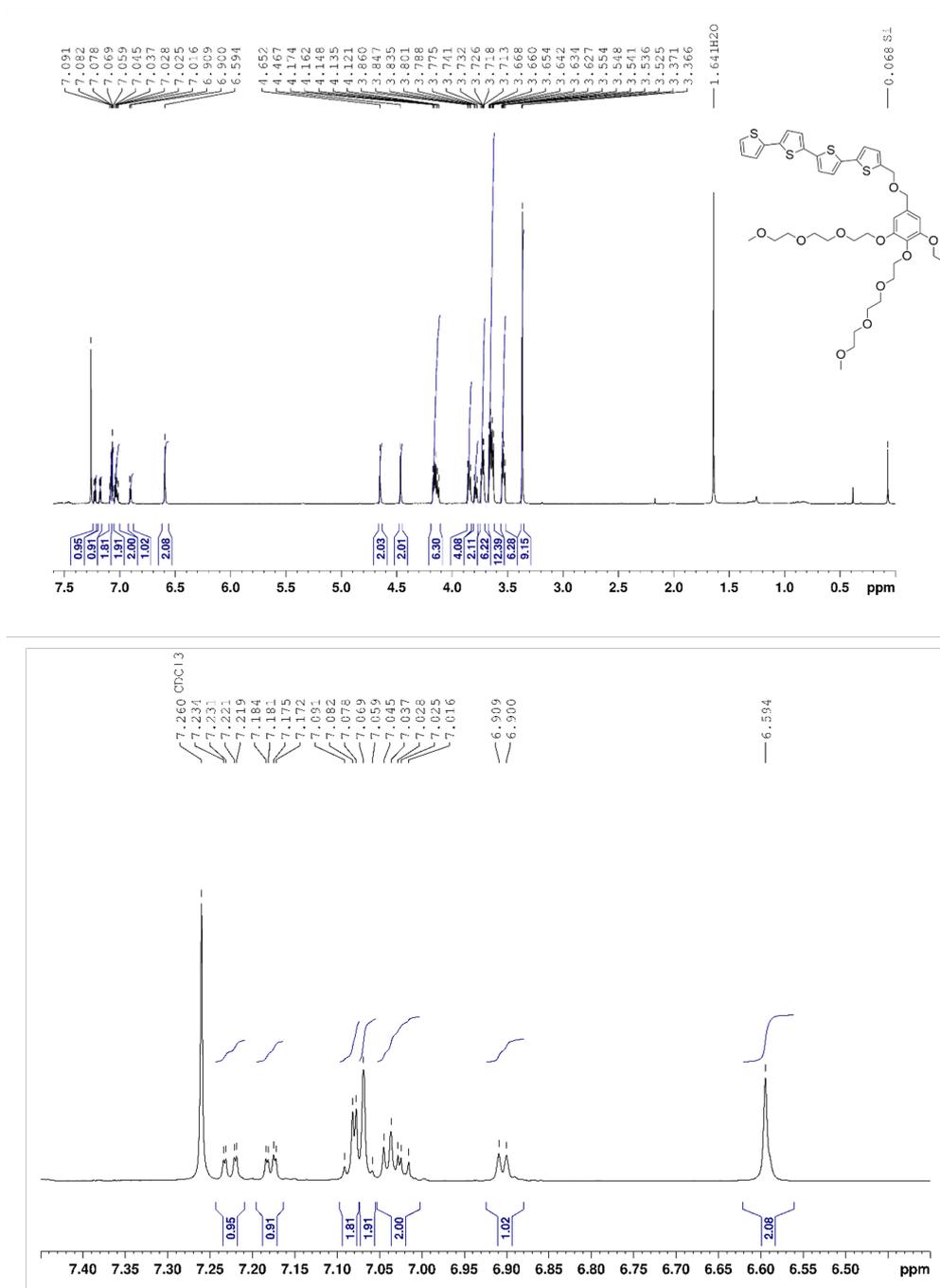

Fig. 12 Full (top) and partial (bottom) $^1$H NMR spectrum (400 MHz, CDCl$_3$) of compound 3

Fig. 13 $^{13}$C$^{1}$H NMR spectrum (101 MHz, CDCl$_3$) of compound 3

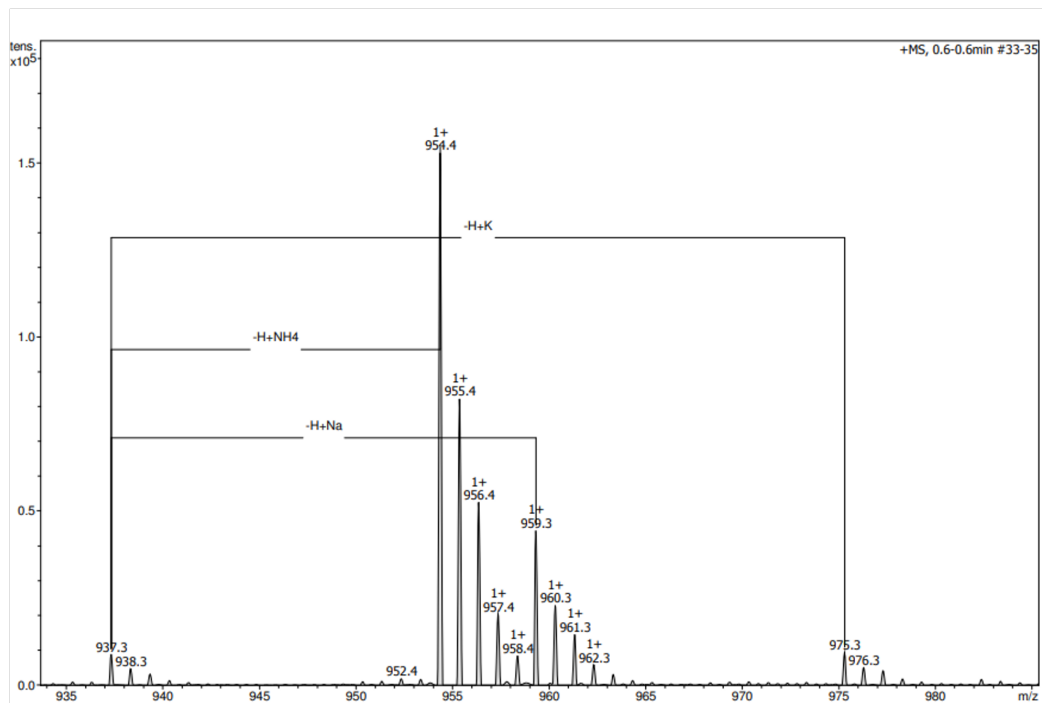

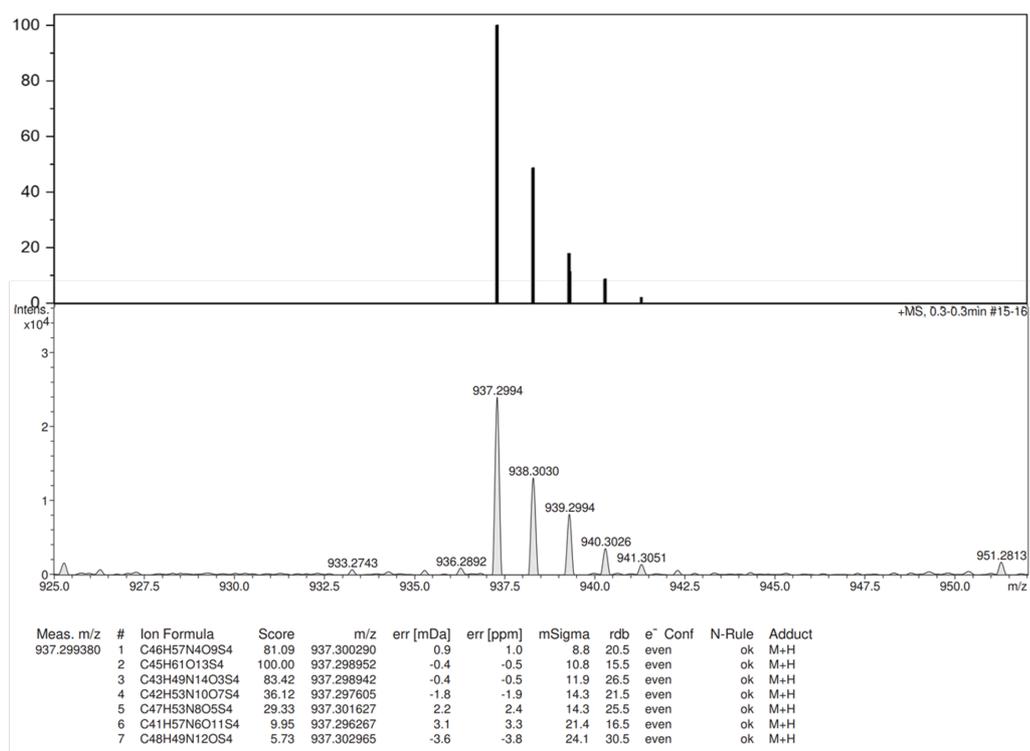

Fig. 14 High-Resolution ESI-TOF (positive mode) mass spectrum (top), theorical and experimental isotopic profiles (bottom) of compound 3